
\magnification 1200
\hsize 15 true cm
\vsize 24 true cm
\parskip 3pt plus 1pt minus 1pt
\count0=0
\nopagenumbers
\rightline{March 22, 1994}
\vskip 6 true cm
\centerline{\bf{On Certain Quantum Deformations of
$gl(N,R)$.}\footnote*{Accepted for publication in JMP.}}
\vskip 1 true cm
\centerline{\bf{David FAIRLIE}}
\vskip 0.5 true cm
\centerline{Department of Mathematical Sciences}
\centerline{University of Durham, Durham, DH1 3LE, England}
\vskip 1 true cm
\centerline{\bf{Jean NUYTS}}
\vskip 0.5 true cm
\centerline{University of Mons, 7000, Mons, Belgium}
\vskip 5 true cm
\noindent Abstract : In this paper all deformations of the
general linear group, subject to certain restrictions which in
particular ensure a smooth passage to the Lie group limit, are
obtained. Representations are given in terms of certains sets of
creation and annihilation operators. These creation and
annihilation operators may belong to a generalisation of the
$q$-quark type or $q$-hadronic type, of $q$-boson or $q$-fermion type.
We are also led to a natural definition of $q$-direct sums of q-algebras.
\vfill\eject
\footline{\hfil\rm{\number\count0}\hfil}
\leftline{1. {\bf{Introduction.}}}
\vskip 0.5 true cm
\par In the last years there have been numerous articles
extending the study of Hopf algebras with their comultiplication
and trying to see if these algebras can be of any use in
certain corners of physical theories. Another related approach
puts less emphasis on the comultiplication, which is sometimes
dropped completely, but insists on realizing deformations of
classical algebras in terms of deformed creation and
annihilation operators.
\par This article, as will be explained shortly in more detail,
belongs to the second approach. One of its possible aims would
be the construction of a new field theory with, obviously, new
types of statistics. We will soon make our
assumptions precise and justify them. But let us emphasise
immediately that, to replace the comultiplication, we will be
led to the new and, we believe, important notion of $q$-direct
sum of representations built out of the deformed creation and
annihilation operators.
\par One of the main characteristic features of operators in
Hilbert spaces, which underly quantum mechanics, is the
associative law of composition. In a series of papers$^{[1,2,3]}$, we have
studied various aspects of quadratic algebras which permit the
reversal of ordering of any pair of operators and have obtained
restrictions on the braiding parameters which arise from the
requirements of associativity.
For example, while normally creation and
annihilation operators belonging to
different particles are assumed to satisfy (deformed) commutation or
anticommutation relations with no central terms, we have investigated in
$^{[2]}$ the possibilty of the addition of central terms,
which we call neutral
operators. As an extension of the same nature
with different closure properties
has recently been investigated by Polychronakos and others\rlap.$^{[4,5,6]}$
The purpose of this article is to
study certain deformations of the general linear group in $N$
dimensions and their realisation in terms of sets of creation
and annihilation operators.
The constraint which we shall impose to limit our search
and which we make more precise below is that as the deformation parameters
approach their undeformed limits
no operator relations drop out of the algebra.
This rules out deformations into superalgebras.
\par The algebra $gl(N,R)$ of the group $GL(N,R)$, i.e. the
group of linear
transformations in $N$-dimensions on the reals,
has $N^2$ generators. We will study its deformations,
which we will call by definition $al(N,R)_q$,
subject to the two restrictions
\medskip
\item{1)} the commutator terms are deformed into quommutators in the
usual basis of $gl(N,R)$.
\item{2)} the structure constants which are zero
in the usual basis for $gl(N,R)$ remain zero for $al(N,R)_q$. Note that this
assumption is not true for the deformation of $SU(n+m)$ into the superalgebra
 $SU(n|m)$.
\medskip
\noindent The generic form of these deformations is made explicit
in (1.1).
\par The $N^2$ generators of $al(N,R)_q$
are denoted by $E(j,k),\ j=1,\ldots,N,\
k=1,\ldots,N$. These operators are supposed to obey a product
law which is in general non commutative but associative.
\par In agreement with the usual commutation relations of
$gl(N,R)$ and subject to the restrictions above, the
general form of the quommutation relations are postulated to be
of the restricted form
$$\eqalign{
\left [E(j,k),E(l,m)\right ]_{p(j,k,l,m)}=
  &\delta_{kl}(1-\delta_{jm})g_1(j,k,m)E(j,m)		\cr
  &-\delta_{mj}(1-\delta_{kl})g_2(l,m,k)E(l,k)		\cr
  &+\delta_{kl}\delta_{mj}(g_3(j,k)E(j,j)-g_4(j,k)E(k,k))   \cr}
\eqno(1.1a)
$$
where the symbol $[X,Y]_{p}$ is a shorthand for
$$
\left [X,Y\right]_{p}=p^{-1}X*Y-p\ Y*X
\eqno(1.1b)
$$
and $p$, $g_1$, $g_2$ , $g_3$ and $g_4$ are indexed sets of
complex numbers
which are supposed to be essentially non-zero.
For the normal $gl(N,R)$ algebra all the parameters $p,\ g_1, \
g_2,\ g_3$  and
$g_4$ are  $1$. If some $p$'s are equal to $i$ we have
anticommutation type rules.
\par On these quommutation relations, the following braiding or
associativity requirements should be imposed
$$
\left (E(j,k)*E(l,m)\right )*E(p,q)=E(j,k)*\left
(E(l,m)*E(p,q)\right )
\eqno(1.2)
$$
where the parentheses specify in which order the products
have to be performed.
\par It is clear that the generalized structure constants in
(1.1) have to fulfill quite a number of conditions resulting
from the associativity and the symmetry requirements.
One way to take these into account is to define a ``normal''
product of the operators $E(j,k)$ for example by
\medskip
\item{-} $E(j,k)$ is on the right of $E(l,m)$ if $l<j$,
\item{-} $E(j,k)$ is on the right of $E(j,m)$ if $m<k$.
\medskip
\par The rule (1.1) then allows the rewriting of any product of
operators as a sum of normal ordered products and the
associativity requirements guarantee that the resulting
expression does not depend on the order in which the required
transpositions are performed.
\par We begin  by a general discussion about what
should be changed for quantum algebras of the form we are
considering as compared to
normal commutation relations customary for Lie algebras
and their representation. In particular the notion underlying
the usual concept of a direct sum of representations will be
discussed and generalized.
\par We shall write explicitly the solutions for the generalized
structure constants for $al(N,R)_q$ for all $N$, and
focus on $al(3,R)_q$ and its realisation
in terms of creation and annihilation operators
forming triplets or nonets, and which satisfy quantum type
quommutation relations. It is suggested that for these
representations the usual notion of coproduct be replaced by a
more natural newly defined $q$-direct sum of representations.
\vfill\eject
\leftline{2. {\bf{General considerations on quantum algebras.}}}
\vskip 0.5 true cm
\par In this section we quote a few results relevant to quantum
algebras and in particular discuss some representations
including matrix representations as well as representations in
terms of quantum creation and annihilation operators of what we
will call boson or fermion type. Some consideration will also be
paid to the all important notion of $q$-direct sum
of representations.
\vskip 0.5 true cm
\leftline{2.a Structure constants. Adjoint representation.}
\vskip 0.5 true cm
\par Suppose we start with a general quantum algebra of the type which
we have been considering in the introduction i.e. of the chosen
form for $i\neq j$
$$
V_i*V_j=q_{ij}V_j*V_i+f_{ij}^{\phantom{ij}k}V_k
\eqno(2.1)
$$
where we will call (see (1.1)) the $q_{ij}\equiv p^2_{ij}$
the quantum parameters and
$f_{ij}^{\phantom{ij}k}\equiv p_{ij}g_{ij}^{\phantom{ij}k}$ the
structure constants. These
parameters satisfy the trivial symmetry requirements always for
$i\neq j$
$$\eqalign{
q_{ij}&={1\over q_{ji}} \ \ \ \ ,  		\cr
f_{ij}^{\phantom{ij}k}&=-q_{ij}f_{ji}^{\phantom{ij}k}\ \ \ \ .		\cr}
\eqno(2.2)
$$
\par Let us remark that these relations could be supplemented by
generalized quommutators of the form
$$
V_i^2=f_{ii}^{\phantom{ii}k}V_k
\eqno(2.3)
$$
which we will not consider here. Another way of stating this
choice is to postulate in (2.1) that
$$\eqalign{
q_{ii}&=1\ \ \ \ ,		\cr
f_{ii}^{\phantom{ii}k}&=0\ \ \ \ . 		\cr}
\eqno(2.4)
$$
\par The associativity requirements
$$
\left ( V_i*V_j\right )*V_k=V_i*\left ( V_j*V_k\right )
\eqno(2.5)
$$
lead to restrictions on the quantum parameters and on the
structure constants.
They replace the Jacobi
identities customary for the structure constants of the Lie
algebras. They are of the form
$$\eqalignno{
(1-q_{ij}q_{ik}q_{mi})f_{jk}^{\phantom{jk}m}&=0&(2.6a) 		\cr
(1-q_{jm}q_{km})f_{jk}^{\phantom{jk}m}&=0   &(2.6b) 		\cr
(1-q_{im})f_{in}^{\phantom{jk}n}
  -(1-q_{in})f_{im}^{\phantom{im}m}&=0  &(2.6c) 		\cr
f_{ij}^{\phantom{ij}p}f_{pk}^{\phantom{pk}m}
   -q_{jk}f_{ik}^{\phantom{ik}p}f_{pj}^{\phantom{pj}m}
   -f_{jk}^{\phantom{jk}p}f_{ip}^{\phantom{ip}m}
   +(q_{ij}-1)f_{ik}^{\phantom{ik}k}f_{jk}^{\phantom{jk}m}&=0
  &(2.6d)\ \ \ \ . 		\cr}
$$
It should be stressed that in (2.6) all the indices appearing
(except the summation index $p$ and the index $m$ in (2.6d)) should
assume different
values. Indeed, though (2.6b) is reducible to (2.6a) when $m=i$
this is not so for (2.6c) which is a weaker condition than the
condition
which would follow from (2.6a) if the indices are allowed to
take equal values (say $m=j$).
\par In view of (2.6d) it is tempting to introduce what we will
call the adjoint
representation of the abstract q-algebras, in analogy with what
is done for Lie algebras. In other words we ask what are the
further restrictions which enable the $N$ matrices $V_i$
whose elements are defined in terms of the structure constants by
$$
(V_i)_j^{\phantom {j}k}=f_{ji}^{\phantom{ji}k}
\eqno(2.7)
$$
to form a representation of the qualgebra.
\par It is easy to prove that these $V$'s satisfy the starting algebra
provided that (2.6d) is replaced by the stronger conditions :
$$
f_{ij}^{\phantom{ij}p}f_{pk}^{\phantom{pk}m}
   -q_{jk}f_{ik}^{\phantom{ik}p}f_{pj}^{\phantom{pj}m}
   -f_{jk}^{\phantom{jk}p}f_{ip}^{\phantom{ip}m}=0  \ \ \ \ .
\eqno(2.8)
$$
To be more explicit these conditions mean that
\medskip
\item{a)}First the two last terms of (2.6d) should not be
present, i.e. for $m\neq n$
$$
(1-q_{im})f_{in}^{\phantom{jk}n}=0\ \ \ \ .
\eqno(2.9)
$$
This is an non trivial extension of (2.6a) when indices are allowed to take
certain equal values.
\item{b)} Moreover these are not the only conditions. Indeed
(2.6d) has been derived for unequal indices
$i\neq j\neq k\neq i$ only and (2.8) has to
hold for all indices.
\medskip
\par If the stronger conditions in $a)$ and $b)$ above do not hold, it seems
that an analogous adjoint representation cannot be defined.
We will see later that for the general $al(2,R)_q$ the
conditions for the existence of the adjoint representation do not
follow from the associativity relations while
for $al(N,R)_q$ for $N>2$ these conditions are
automatically satisfied. This will be the essential point
allowing the representation of these q-algebras in terms of
creation and annihilation operators
of $q$-boson or $q$-fermion type. On the other hand we shall
prove that the existence of the requirement of the existence of
the adjoint representation for $al(2,R)_q$ reduces in an essential
way the allowed values of the quantum parameters.
\vskip 0.5 true cm
\leftline{2.b $q$-direct sums of q-algebras.}
\vskip 0.5 true cm
\par  A generalized notion of $q$-direct sum of
isomorphic q-algebras can also be defined as follows.
\par Let $V_i$ and $W_i$ two isomorphic q-algebras with the same
quantum parameters and structure constants of the form (2.1).
We will define the $Z_i$ operators of the $q$-direct sum of the
two algebras by
$$
Z_i=V_i+W_i
\eqno(2.10a)
$$
provided that $V_i$ and $W_j$ satisfy the quantum relations
$$
V_i*W_j=q_{ij}W_j*V_i   \ \ \ \ .
\eqno(2.10b)
$$
\par With these definitions and conditions it is obvious that
the $Z_i$ satisfy the same qualgebra as $V_i$ and $W_i$ without
any further condition. It should be stressed that within a
qualgebra structure (2.10b) is a more natural condition than
simple commutation and that it
reduces to simple commutation in the Lie algebra case, as it should.
\par With this notion we can now define $q$-direct sums of
representations. Contrary to the Lie case where the direct sum
can always be defined, using direct products with the unit matrix,
this is not the case here. We shall however
show that for representations built out of creation and
annihilation operators this definition is natural.
\vskip 0.5 true cm
\leftline{3. {\bf{The qualgebra $al(2,R)_q$.}}}
\vskip 0.5 true cm
\par If we limits ourselves to the case of $al(2,R)_q$, the result of the
braiding relations leads to the general deformation, with a
suitable rescaling of the diagonal operators $E(k,k)$ given by
$$\eqalign{
E(1,1)*E(2,2)&=E(2,2)*E(1,1)  		\cr
E(1,1)*E(1,2)&=\alpha^2 E(1,2)*E(1,1)+\alpha E(1,2)  		\cr
E(2,1)*E(1,1)&=\alpha^2 E(1,1)*E(2,1)+\alpha E(2,1)  		\cr
E(2,2)*E(1,2)&=\beta^2 E(1,2)*E(2,2)+\beta E(1,2)  		\cr
E(2,1)*E(2,2)&=\beta^2 E(2,2)*E(2,1)+\beta E(2,1)  		\cr
E(2,1)*E(1,2)&=\gamma^2 E(1,2)*E(2,1)
              +\gamma\lambda E(1,1)+\gamma\mu E(2,2) \ \ \ \ ,\cr}
\eqno(3.1)
$$
or using
$$
V_{i+2(j-1)}=E(i,j) \ \ \ \ ,
\eqno(3.2)
$$
this is written as
$$\eqalign{
&[V_1,V_2]_{1\over\alpha}=-V_2 		\cr
&[V_1,V_3]_{\alpha}=V_3  		\cr
&[V_1,V_4]=0  		\cr
&[V_2,V_3]_{\gamma}=\lambda V_1+\mu V_4 		\cr
&[V_2,V_4]_{\beta}=V_2   		\cr
&[V_3,V_4]_{1\over\beta}=-V_3  		\cr}
\eqno(3.3)
$$
in the  notation of (2.1).
\par We see that that four meaningful parameters survive
since one of the five parameters above can be rescaled away
(say $\lambda=1$) by
multiplying (rescaling) $E(1,2)$ by a well-chosen factor (or
equivalently rescaling $E(2,1)$).
\par The conditions for the existence of the adjoint
representation restrict drastically the number of free
parameters. Indeed this implies
$$\eqalign{
\alpha^2 =\beta^2 &=1 		\cr
(\gamma^2 -1)(\alpha \mu+\beta\lambda)&=0 \ \ \ \ .		\cr}
\eqno(3.4)
$$
\par Representations of
certain if these q-algebras can be constructed using sets of
$N$ creation operators $a^{\dag}_i,i=1,\ldots,N$ and $N$
annihilation operators $a_i,i=1,\ldots,N$. These operators satisfy
$$\eqalign{
a_ia_j&=x_{ij}a_ja_i   		\cr
a^{\dag}_ia^{\dag}_j&=x_{ij}a^{\dag}_ja^{\dag}_i   		\cr
a_ia^{\dag}_j&=x_{ji}a^{\dag}_ja_i            \cr
a_ia^{\dag}_i&=y_{i}a^{\dag}_ia_i+1  		\cr
   }
\eqalign{
  {\rm{for\ \ }}  i\neq j 		\cr
   }
\eqno(3.5a)
$$
and, as usual,
$$
x_{ij}={1\over x_{ji}}  \ \ \ \ .
\eqno(3.5b)
$$
This set is associative as can be checked easily.
The $a^{\dag}_i$ can be interpreted as the conjugates of the $a_i$
if the $x_{ij}$ are chosen to be of modulus one.
For the case at hand and for $N=2$, the natural operators
$$
E(i,j)=z_{ij}a^{\dag}_i a_j
\eqno(3.6)
$$
satisfy (3.1) provided that the following relations hold
$$\eqalign {
\gamma^2&=\alpha^2\beta^2   		\cr
\lambda&=\alpha\beta\mu		\cr
z_{1,1}&=\alpha		\cr
z_{2,2}&=-{1\over\beta}   		\cr
z_{1,2}z_{2,1}&=-{\mu\gamma\over\beta}  		\cr
y_1&=\alpha^2   		\cr
y_2&={1\over\beta^2}\ \ \ \ .		\cr}
\eqno(3.7)
$$
\par With four creation and annihilation operators satisfying
the same equations (3.5) for $N=4$
we may try to construct
$$
V_i=\sum_{j=1}^{4}\sum_{k=1}^{4}g_{ji}^{\phantom{ij}k}a^{\dag}_ja_k
\ \ \ \ ,
\eqno(3.8)
$$
where the $g_{ji}^{\phantom{ij}k}$'s are non zero when the corresponding
$f_{ji}^{\phantom{ij}k}$'s do not vanish but also for non zero
$g_{11}^{\phantom{11}1}$, $g_{14}^{\phantom{14}4}$,
$g_{41}^{\phantom{41}1}$  and $g_{44}^{\phantom{44}4}$ in order
to allow for a singlet.
The result is more constrained than for two operators.
In fact these cases are not interesting :
the algebra involves commutators and/or anticommutators only.
\vskip 0.5 true cm
\leftline{4. {\bf{The qualgebra $al(3,R)_q$.}}}
\vskip 0.5 true cm
\par Appplying the same technique, one can easily find the most
general $al(3,R)_q$. This algebra can be represented in
terms of creation and annihilation operators either of triplets
of $q$-quarks or of nonets of $q$-fermions or $q$-bosons.
\vskip 0.5 true cm
\leftline{4.a The quommutation relations of $al(3,R)_q$}
\vskip 0.5 true cm
\par It is remarkable that the very presence of the third
dimension reduces drastically the number of free parameters to
just one $q$, once arbitrary rescalings have been taken into account.
\par As for the $al(2,R)_q$ case, it is again convenient to rename the
$E(i,j),i,j=1,2,3$ operators
as $V(\alpha),\alpha=1,\ldots,9$ in the following way
$$
V_{i+3(j-1)}=E(i,j)\ \ \ \ .
\eqno(4.1)
$$
The general quommutation relation are explicitly given in
Appendix A in the form (1.1).
\par It should be stressed that this algebra has only one free
parameter once the rescalings ($V'_i=\rho_iV_i$, no summation over
$i$) have been taken into account. These rescalings do not affect
the quantum parameters but renormalize the structure constants.
The particular rescalings where $\rho_2$ and $\rho_6$ are
arbitrary while
$\rho_1=\rho_5=\rho_9=1,\rho_3=\rho_2\rho_6,\rho_4=\rho_2^{-1},
\rho_7=\rho_3^{-1},\rho_8=\rho_6^{-1}$ leaves the algebra completely
invariant.
\par Since $V_d=V_1+V_5+V_9$ commutes with all the $V$'s, the algebra
$al(3,r)_q$ can be reduced to the direct sum of the algebra of $sl(3,r)_q$
with eight generators and the algebra $u(1)$ with generator $V_d$.
\par The amusing point about these results is that $al(3,r)_q$ (resp.
$sl(3,r)_q$)
possesses three undeformed $gl(2,r)$ (resp. $sl(2,r)_q$) subalgebras,
formerly called, for
$U(3)$, the $i$-spin $(V_1,V_2,V_4,V_5)$, the $u$-spin
$(V_1,V_3,V_7,V_9)$ and the $v$-spin \break
$(V_5,V_6,V_8,V_9)$ . These three
subalgebras have
normal Lie algebra commutation relations. The only place where the quantum
parameter $q$ shows up is in the interplay between the three
unbroken $gl(2,r)$.
\par It should also be stressed that no diagonal $so(3)$ (or
$so(3)_q$) algebra (the algebra generated by say
$\lambda_2,\lambda_5,\lambda_7$ in Gell-Mann's notation) with three
generators can be constructed within the
$V_2,V_3,V_4,V_6,V_7,V_8$ subspace  as a
subalgebra of $al(3,R)_q$ (except when $q=1$).
\par The associativity relations for $al(3,R)_q$ belong
automatically to the strong case such that the adjoint
representation (2.7) holds.
\par The quommutation relation of $al(3,R)_q$ can be obtained
starting from suitable sets of creation and annihilation
operators. In the following we present the results for triplets
or octets of boson or fermion type operators obeying
 commutation relations of quantum type.
\vskip 0.5 true cm
\leftline{4.b Representation of $al(3,R)_q$ with $q$-quark triplets.}
\vskip 0.5 true cm
\par Let us take one collection of annihilation
operators
$a_i, i=1,2,3$ and their corresponding creation operators
$a^{\dag}_i, i=1,2,3$ and
impose upon them
the generic relations (3.5a).
The
$a^{\dag}_i$ can be interpreted as the hermitian conjugates of the $a_i$
if the $x_{ij}$ are chosen to be of modulus one.
If the $x_{ij}$ are not of modulus one we have to define an
involution $\dag$ between the creation and annihilation operators. For
this involution the conjugates of
products of two operators have to be redefined and,
for example, the adjoints of products of two operators can be
defined in the following way
$$\eqalign{
(a_ia_j)^{\dag}&=m_{ij}a_j^{\dag}a_i^{\dag}   		\cr
(a_i^{\dag}a_j^{\dag})^{\dag}&=m_{ij}a_j a_i   		\cr
(a_ia_j^{\dag})^{\dag}&=m_{ji}a_ja_i^{\dag}   		\cr
(a_i^{\dag}a_j)^{\dag}&=m_{ji}a_j^{\dag} a_i   		\cr}
\eqno(4.2)
$$
with $m_{ij}$ real given by
$$
m_{ij}=\sqrt{x_{ij} x_{ij}^*}\ \ \ \ .
\eqno(4.3)
$$
\par We now construct, in a natural way, the $al(3,R)_q$
operators by
$$
E(i,j)=z_{ij}a^{\dag}_i a_j \ \ \ \ .
\eqno(4.4)
$$
\par Inserting this ans\"atz into (A.2) we find that there are
two solutions for which (4.4) is a
representation.
\par Either
$$\eqalign{
y_i&=1\ \ \ \ \ \ \ \ \ i=1,2,3  		\cr
x_{12}x_{23}x_{31}&=p^2	\ \ \ \ ,	\cr}
\eqno(4.4a)
$$
\par or
$$\eqalign{
y_i&=-1\ \ \ \ \ \ \ \ \ i=1,2,3   		\cr
x_{12}x_{23}x_{31}&=-p^2   		\cr
a_i^2&=0		\cr
a_i^{\dag2}&=0  \ \ \ \ . 		\cr}
\eqno(4.4b)
$$
\par In the first solution the quarks are of $q$-bosonic type
while in the second
case they are of $q$-fermionic type.
\par Moreover the $z$'s
have to be chosen as follows
$$\eqalign{
z_{ii}&=1\ \ \ \ \ \ \ \ \ i=1,2,3 		\cr
z_{ij}&={1\over z_{ji}} \ \ \ \ \ \ \ \ i\neq j 		\cr
z_{12}z_{23}z_{31}&=p 	\ \ \ \ .	\cr}
\eqno(4.4c)
$$
\par At this point it is perhaps useful to remark that, for
$p=1$, the usual Lie algebra of $GL(3,R)$ can be represented
with $q$-quarks i.e. with creation and annihilation operators
which are not of pure bosonic or fermionic nature.
\par From this representation another (the conjugate
representation) can be constructed as follows
$$
\hat E(i,j)=-\hat z_{ij}a^{\dag}_j a_i
\eqno(4.5a)
$$
where
$$\eqalign{
\hat z_{ii}&=1\ \ \ \ \ i=1,2,3 		\cr
\hat z_{ij}&={1\over \hat z_{ji}}		\cr
\hat z_{12}\hat z_{23}\hat z_{31}&={1\over p} \ \ \ \ .   		\cr}
\eqno(4.5b)
$$
It is easily checked that the $E(i,j)$'s and the $\hat E(i,j)$'s
satisfy the quommutation relations (4.2). These results are consistent
with the usual relation between a representation and its
conjugate $\hat E(i,j)=-E^{\dag}(i,j)$ if the adjoint is taken
as the involution defined above (see (4.2-3).
\par We now make a very important point. Suppose that
we have a second set of creation and annihilation operators
$b_i,b^{\dag}_i,i=1,2,3$ which satisfy exactly the same
quommutation relations as the $a,a^{\dag}$ and hence with the
obvious definitions, the $F$'s
$$
F(i,j)=z_{ij}b^{\dag}_i b_j
\eqno(4.6)
$$
satisfy the same $al(3,R)_q$ algebra.
\par Let us we try to impose between the $a$'s and the $b$
quommutation relations of the general form
$$\eqalign{
a_ib_j&=\bar\alpha_{ij}\ b_ja_i 		\cr
a_i^{\dag}b_j^{\dag}&=\alpha_{ij}\ b_j^{\dag}a_i^{\dag}  		\cr
a_ib_j^{\dag}&=\bar\beta_{ij}\ b_j^{\dag}a_i   		\cr
a_i^{\dag}b_j&=\beta_{ij}\ b_ja_i^{\dag}		\cr
   }
\eqno(4.7)
$$
where the indices obviously run from 1 to 3.
\par The naive sums of the operators $E$ and $F$
$$
G(i,j)=E(i,j)+F(i,j)
\eqno(4.8)
$$
satisfy
the same $al(3,R)_q$ qualgebra relations if the following
restrictions upon the braiding relations (4.7) hold.
\medskip
\item{1)} The diagonal elements $\alpha_{ii},\beta_{ii},
\bar\alpha_{ii},\bar\beta_{ii}$ are essentially free parameters
except that
$$
\alpha_{ii}\beta_{ii}\bar\alpha_{ii}\bar\beta_{ii}=1\ \ \ \ \ i=1,2,3
\ \ \ \ .
\eqno(4.9a)
$$
\item{2)} The three first sets are related to the fourth by
$$\eqalign{
\bar\alpha_{ij}&={1\over\bar\beta_{ii}\beta_{jj}}{1\over\alpha_{ji}}	\cr
\bar\beta_{ij}&={\bar\beta_{ii}\over\alpha_{jj}}{\alpha_{ji}}\ \
\ \ \ \ i\neq j  		\cr
\beta_{ij}&={\beta_{jj}\over\alpha_{ii}}{\alpha_{ji}}   		\cr}
\ \ \ \ .
\eqno(4.9b)
$$
\item{3)} Within the fourth set the following relations hold
$$\eqalign{
\alpha_{ij}&={\alpha_{ii}\alpha_{jj}}{1\over\alpha_{ji}}\ \ \ i>j 	\cr
\alpha_{23}&={p^2\alpha_{22}\alpha_{13}\over
\alpha_{12}} 		\cr}
\ \ \ \ .
\eqno(4.9c)
$$
\medskip
\par In total there are thus nine parameters of a diagonal
type and two parameters $\alpha_{12},\alpha_{13}$ of a
non-diagonal type. A particularly simple solution, the
interesting one in fact, is obtained when all the free diagonal
elements are chosen to be equal to $y_1$ i.e. +1 or -1 when the
triplets are boson-like or fermion-like respectively while the
$\alpha$'s are identified with the $x$'s
$$\eqalign{
\alpha_{ij}&=\bar\alpha_{ij}=\beta_{ji}=\bar\beta_{ji}=x_{ij}
\ \ \ \ \ \ \ \ \ \ \ \ \ \ i\neq j 		\cr
\alpha_{ii}&=\beta_{ii}=\bar\alpha_{ii}=\bar\beta_{ii}=y_1 \ \ \ \ .\cr
   }
\eqno(4.10)
$$
\par For this solution the quommutation relations between two
copies of creation and annihilation operators become
$$\eqalign{
a_ib_j&=x_{ij}\ b_ja_i 		\cr
a_i^{\dag}b_j^{\dag}&=x_{ij}\ b_j^{\dag}a_i^{\dag}  		\cr
a_ib_j^{\dag}&=x_{ji}\ b_j^{\dag}a_i   		\cr
a_i^{\dag}b_j&=x_{ji}\ b_ja_i^{\dag}\ \ \ \ .		\cr
   }
\eqno(4.11)
$$
\par One might hope that this remark would allow the
construction of a $q$-field theory with quommutation relations.
This appears to be not the case however since, when a field of a
given $i$-index at one space time point is
constructed as a sum of creation and annihilation operators,
there is no overall quommutation relation with another field of
$j$-index at a different space time point unless $x_{ij}^2=1$.
This is a specific illustration of a difficulty common to the
problem of constructing a coherent field theory out of $q$-operators.
\vskip 0.5 true cm
\leftline{4.c Representation of $al(3,R)_q$ with $q$-quarks nonets.}
\vskip 0.5 true cm
\par In a completely analogous way the qualgebra $al(3,R)_q$ can
be represented using nonets of $q$-quarks. Without going into the
details let us write explicitly one such solution, using the
$q$'s and the $f$'s obtained from (A.2) once the corresponding
equations have been written in the form (2.1)
$$
V_i=f_{ji}^{\phantom{ji}k}a^{\dag}_ja_k
\eqno(4.12)
$$
where the quommutation relations of the nine creation and
nine annihilation operators follow (4.3) with the indices
running from 1 to 9. All the $y_i$ are equal to $y_1$ of square 1
and
$$
x_{ij}=y_1q_{ij}{m_i\over m_j}
\eqno(4.13)
$$
where the $m$'s are arbitrary factors present even when $p=1$.
The square of the creation and annihilation operators have
to be equal to zero if $y_1=-1$, i.e. in the case of nonets
of $q$-operators of the fermion type.
\par The arbitrariness for the braiding coefficients $x_{ij}$ as
seen in (4.4) and (4.13) demonstrate that even the
usual Lie algebras (i.e. when the $q_{ij}$'s are equal to one)
can be represented with operators which do
not fully commute nor anticommute.
\vskip 10pt
\leftline{5. {\bf{The q-algebras $al(N,R)_q$.}}}
\vskip 0.5 true cm
\par A canonical form of the general qualgebra
$al(N,R)_q$ for any $N$ can easily be constructed along the same
lines as we have done for $al(3,R)$ and under the same
restrictions as defined in the introduction. Let us present it in the
notation of equation (1.1). The parameters $p(j,k,l,m)$
can be constructed from a set of arbitrary parameters $q$.
\par Indeed let $q_{ij}$ with $i<j=1,\ldots,N$ be an arbitrary set of
complex numbers and let
$$\eqalign{
q_{1j}&=1 \ \ \ \ \ j=2,\ldots,n		\cr
q_{jj}&=1 \ \ \ \ \ j=1,\ldots,n		\cr
q_{ij}&={1\over q_{ji}}\ \ \ \ \ \ i>j	\ \ \ \ .	\cr}
\eqno(5.1a)
$$
There are thus $N(N-1)/2$ arbitrary complex parameters $q$.
We then obtain
$$
p(j,k,l,m)={q_{jl}q_{km}\over q_{kl}q_{jm}} \ \ \ \ .
\eqno(5.1b)
$$
Moreover the right hand side in (1,1) differs very little from
those of the
underformed $gl(N,R)$ algebra. Indeed, one can chose a specific
normalisation of the
$V_i$'s such that
$g_1$, $g_2$, $g_3$ and $g_4$ all retain their undeformed value.
\par The q-algebras $al(N,R)$ were already
written down$^{[7]}$ in a slightly different form, starting from $N$ creation
and $N$ annihilation operators. We would like to stress here that,
except for renormalisations of the
operators $V_i$, this is the most general solution compatible with our
starting hypothesis.
\par It is easy to see that representation of these operators can be
written in terms of a set of basic $N$ creation and $N$ annihilation
operators but also in terms of $N^2$ creation and $N^2$ annihilation
operators by generalising in an obvious fashion what we have done for
$al(N,3)_q$.
\vskip 0.5 true cm
\leftline{6. {\bf{Conclusions.}}}
\vskip 0.5 true cm
\par One objective of this investigation was to find all the
deformations of $gl(N,R)$ subject to natural restrictions. We
succeeded and were led, in the process, to realize the deformed
algebras in terms of deformed sets of creation and annihilation
operators and to introduce a generalisation of the direct sum of
algebras and of their representations.
In his studies of braided tensor categories Majid has applied a
similar concept to the $q$-direct sum to define an addition law
for momentum in a braided Lorentz group$^{[8]}$.
\par In order to proceed further in the direction of application
of these ideas and results to physics$^{[9]}$, it would seem necessary to
construct local fields out of the creation and annihilation
operators. These fields should satisfy generalized canonical
commutation relations.
Unfortunately the rules for quommuting different sets of copies
of the operators corresponding to different momenta or
different space-time points turn out be be incompatible with a
naive extension in which fields are linear combinations
involving both creation and annihilation operators. Hence the
construction of a genuine extension of the usual field theory
remains a problem.
\vfill\eject

\vfill\eject
\leftline{References}
\vskip 1 true cm
\item{[1]} Fairlie  D.B. and  Nuyts J.,
{\it Jour. Phys.} A {\bf{24}}, L1001 (1991).
\item{[2]} Fairlie  D.B. and  Nuyts J.,
{\it Zeitschrift f\"ur Physik} {\bf{56}}, 237 (1992).
\item{[3]} Fairlie  D.B. and  Nuyts J.,
{\it Jour. Math. Phys.} {\bf{34}}, 4441 (1993).
\item{[4]} Polychronakos A.P.,
{\it Phys. Rev. Lett.}, {\bf{69}}, 703 (1992).
\item{[5]} Brink L.,  Hansson T. and  Vasiliev M.A.,
{\it Phys. Lett.}, {\bf{B286}}, 109 (1992).
\item{[6]} Turbiner A.T. and  Post G. Operator identities, representations of
algebras and the problem of normal
ordering. CERN preprint CRN-93/53 (1993).
\item{[7]} Fairlie D.B. and  Zachos C.K.,
{\it Phys.Lett.} {\bf{256B}}, 43 (1991).
\item{[8]} Majid S., {\it Jour. Math. Phys.}\ {\bf
34}, 2045 (1993).
\item{[9]} Greenberg O.W., {\it Phys. Rev. Lett.}\ {\bf
{64}}, 705 (1990).
\vfill\eject

\leftline{Appendix A. The algebra al(3,R).}
\par With suitable renormalisations of the operators $V_i$ by
coefficients $\rho_i$ the most general $al(3,R)$, obtained by
deformations of $gl(3,R)$ subject to the conditions written in
the first paragraph of the introduction, is given by the
set of quommutators given in (A.1) below.
\par By simply looking at (A.1) it is easy to see that the three
al(3,R) subalgebras with
the three sets of generators
$\{V_1,V_2,V_4,V_5\}$, $\{V_1,V_3,V_7,V_9\}$ and
$\{V_5,V_6,V_8,V_9\}$ are undeformed corresponding to i-spin,
u-spin and v-spin.
On the other hand an $o(3)$, deformed or not, built out of three
linear combinations of the generators
$\{V_2,V_3,V_4,V_6,V_7,V_8\}$, cannot be constructed.
$$\eqalign{
&[V_1,V_2]=-V_2   		\cr
&[V_1,V_3]=-V_3   		\cr
&[V_1,V_4]=V_4		\cr
&[V_1,V_5]=0  		\cr
&[V_1,V_6]=0  		\cr
&[V_1,V_7]=V_7		\cr
&[V_1,V_8]=0  		\cr
&[V_1,V_9]=0  		\cr
&[V_2,V_3]_{p}=0  		\cr
&[V_2,V_4]=V_5-V_1		\cr
&[V_2,V_5]=-V_2   		\cr
&[V_2,V_6]_{p}=- V_3   		\cr
}\quad\quad
\eqalign{
&[V_2,V_7]_{1\over p}= V_8 		\cr
&[V_2,V_8]_{1\over p}=0		\cr
&[V_2,V_9]=0  		\cr
&[V_3,V_4]_{p}=V_6		\cr
&[V_3,V_5]=0  		\cr
&[V_3,V_6]_{p}=0  		\cr
&[V_3,V_7]=V_9-V_1		\cr
&[V_3,V_8]_{1\over p}=- V_2   		\cr
&[V_3,V_9]=-V_3   		\cr
&[V_4,V_5]=V_4		\cr
&[V_4,V_6]_{1\over p}=0   		\cr
&[V_4,V_7]_{p}=0  		\cr
}\quad\quad
\eqalign{
&[V_4,V_8]_{p}=V_7		\cr
&[V_4,V_9]=0  		\cr
&[V_5,V_6]=-V_6   		\cr
&[V_5,V_7]=0  		\cr
&[V_5,V_8]=V_8		\cr
&[V_5,V_9]=0  		\cr
&[V_6,V_7]_{p}=- V_4  		\cr
&[V_6,V_8]=V_9-V_5		\cr
&[V_6,V_9]=-V_6   		\cr
&[V_7,V_8]_{p}=0  		\cr
&[V_7,V_9]=V_7		\cr
&[V_8,V_9]=V_8		\cr
                                       }
\eqno(A.1)
$$
\par If $p$ is chosen to be $+1$ we have the undeformed
$gl(3,R)$. If $p$ is chosen to be $+i$ and $\rho_3=-\rho_7=i$
one obtains an algebra with commutators and anticommutators
only. The anticommutators are specifically given in (A.2).
$$\eqalign{
&[V_2,V_3]_{+}=0 		\cr
&[V_2,V_6]_{+}=- V_3 		\cr
&[V_2,V_7]_{+}=- V_8		\cr
&[V_2,V_8]_{+}=0  		\cr
&[V_3,V_4]_{+}=-V_6  		\cr
&[V_3,V_6]_{+}=0 		\cr
&[V_3,V_8]_{+}=- V_2  		\cr
&[V_4,V_6]_{+}=0  		\cr
&[V_4,V_7]_{+}=0 		\cr
&[V_4,V_8]_{+}=-V_7  		\cr
&[V_6,V_7]_{+}=- V_4 		\cr
&[V_7,V_8]_{+}=0 		\cr
                                      }
\eqno(A.2)
$$
\vfill
\end